\documentclass[12pt]{article}
\usepackage{amsfonts}
\usepackage{amsmath}
\usepackage{amssymb}
\usepackage{fancyheadings}
\usepackage{latexsym}
\usepackage{citesort}

\newcommand{\rf}[1]{(\ref{#1})}

\newcommand{\pt}{\partial}

\def\a{\alpha}
\def\b{\beta}
\def\g{\gamma}

\def\d{\delta}

\def\l{\lambda}
\def\m{\mu}
\def\n{\nu}
\def\o{\omega}

\def\s{\sigma}

\def\cN{{\cal N}}

\newcommand{\be}{\begin{equation}}
\newcommand{\ee}{\end{equation}}

\newcommand{\bea}{\begin{eqnarray}}
\newcommand{\eea}{\end{eqnarray}}
\newcommand{\ft}[2]{{\textstyle\frac{#1}{#2}}}
\def\fft#1#2{{\frac{#1}{#2}}}
\def\ul#1{\underline{#1}}
\newcommand{\half}{\frac{1}{2}}

\def\rme{{\rm e}}
\def\rmi{{\rm i}}
\def\rmd{{\rm d}}

\begin{document}

\begin{titlepage}

\font\cmss=cmss10 \font\cmsss=cmss10 at 7pt
\leftline{\tt hep-th/0701115}

\vskip -0.5cm \rightline{\small{\tt KUL-TF-07/04}}

\vskip .7 cm

\hfill
\vspace{18pt}
\begin{center}
{\Large \textbf{The Stability of D-term Cosmic Strings}}
\end{center}

\vspace{6pt}
\begin{center}
{\large\textsl{Andr{\'e}s Collinucci, Paul Smyth and Antoine Van Proeyen}}

\vspace{25pt}
\textit{Institute for Theoretical Physics, K.U. Leuven,\\ Celestijnenlaan 200D, B-3001 Leuven, Belgium}\\
\end{center}

\vspace{12pt}

\begin{center}
\textbf{Abstract}
\end{center}

In this note, which is based on hep-th/0611111, we review the stability of the static, positive deficit angle D-term string solutions of $D=4~,~\cN=1 $ supergravity with a constant Fayet-Iliopoulos term. We prove the semi-classical stability of this class of solutions using standard positive energy theorem techniques. In particular, we discuss how the negative deficit angle D-term string, which also solves the Killing spinor equations, violates the dominant energy condition and so is excluded from our arguments. 

\vspace{4pt} {\small \noindent

}

\vfill

\noindent\small{\textit{This is the transcript of a talk given by PS at the RTN project `Constituents, Fundamental Forces and Symmetries of the Universe' conference, Napoli, October 9- 13, 2006.}}

\vskip 5.mm
\hrule width 5.cm
\vskip 2.mm
{\small
\noindent e-mail: andres.collinucci, paul.smyth, antoine.vanproeyen@fys.kuleuven.be}

\end{titlepage}


\section*{Introduction}

There has been a remarkable resurgence of interest in cosmic strings in recent years (see \cite{rr}, for instance). An observation of the double galaxy gravitational lens candidate CSL-1 \cite{Sazhin:2003cp} which was indicative of a cosmic string sparked particular interest in the field. Further observations of CSL-1 by the Hubble space telescope later showed that this was in fact a pair of very similar, but distinct galaxies \cite{Agol:2006fb}. Despite this, the interest in cosmic strings remains high, and other, indirect, experimental evidence is suggestive of the existence of these objects \cite{Schild:2004uv}. 

There have also been considerable theoretical advances in our understanding of supersymmetric cosmic strings. String topological defects were found some time ago in globally supersymmetric theories \cite{Davis:1997bs}, but were only recently embedded into supergravity \cite{dkvp}. The particular class of local cosmic strings we are interested in can be found as solitonic solutions supported by a $D$-term potential in four-dimensional  $\cN=1$ supergravity with constant Fayet-Iliopoulos terms \cite{dkvp}. A $D$-term string can also be understood as a D$_{1+q}$-brane wrapping a calibrated $q$-cycle in an internal manifold of a string theory compactification \cite{dkvp}. In fact, these $D$-term string solutions were found previously as point-like solutions in three-dimensional supergravity \cite{3d}. 

Much attention has focused on cosmological aspects of string theory cosmic strings, e.g. string networks \cite{cmp}. However, it is surprising to note that the stability of a single, isolated
supersymmetric string solution of supergravity has not been discussed. Bogomol'nyi bounds for general cosmic strings were constructed originally by Comtet and Gibbons \cite{Comtet:1987wi} and the energy of local string solutions in current discussions, including the $D$-term strings,
is usually defined using such Bogomol'nyi-type arguments \cite{dkvp,Achucarro:2005vz}. However, as noted in \cite{Achucarro:2005vz}, a Bogomol'nyi bound does not prove the stability of such local string solutions, as one is implicitly assuming that the solutions remain axisymmetric. It is therefore possible that non-axisymmetric perturbations or string worldvolume perturbations could lead to instabilities.

In this article, which is a summary of \cite{csvp}, we discuss the semi-classical stability of the $D$-term string solution of $D=4~,~\cN=1 $ supergravity with a constant Fayet-Iliopoulos term. Regardless of the particular theory in which one is interested, the stability of cosmic strings is necessary if we hope to observe them.  We apply the spinorial Witten-Nester method to prove a positive energy theorem for the D-term cosmic string background with positive deficit angle. We also pay particular attention to the negative deficit angle $D$-term string, which is known to violate the dominant energy condition. Within the class of string solutions we consider, this violation implies that the negative deficit angle $D$-term string must have a naked pathology and therefore the positive energy theorem we prove does not apply to it.

\section*{The $D$-term string in $\cN=1$ supergravity}

Let us begin by briefly reviewing the relevant aspects of four-dimensional $\cN=1$ supergravity with constant Fayet-Iliopoulos terms \cite{dkvp}. The Lagrangian for the bosonic sector of this theory is\footnote{We are using natural units, setting $M_P=1$. We direct the reader to \cite{csvp} for a full explanation of our conventions and a complete list of references.},
\begin{eqnarray}
e^{-1}{\cal L}=\frac{1}{2}R -\hat {\partial }_\mu
\phi\, \hat {\partial }^\mu \phi^*
 -\ft{1}{4} F_{\mu \nu } F^{\mu \nu }- \frac{1}{2} D^2\,,
 \label{bosonic2}
\end{eqnarray}
where $\phi$ is the $U(1)$-charged Higgs field, the K{\"a}hler potential is given by $K=\phi^*\phi$ and the superpotential vanishes. The $D$-term  potential is defined by $ D\, = g\xi - g \phi^*\, \phi $, where $\xi$ is a constant that we choose to be positive. $W_\mu$ is an abelian gauge field and we define
\begin{equation}
F_{\mu\nu}\equiv {\partial }_\mu W_\nu- {\partial }_\nu W_\mu\, ,\qquad
\hat {\partial }_\mu \phi\equiv ({\partial }_\mu -\rmi g W_\mu) \phi\,.
\end{equation}
The fermions are Majorana spinors, however it is often convenient to
split them into complex parts using left and right projectors $ P_{L,R} =
\half(1\pm\g_5)$. The supersymmetry
transformations for the fermions (the Killing spinor equations) can then
be written as \bea \delta \psi _{\mu}  &=& \hat\nabla_\mu \epsilon =
\nabla_\mu\epsilon + \frac{\rmi}{2} \g_5 A_\mu^B \epsilon\,,
\label{susy1} \\
 \delta \chi_L &=&  \half(\not\! {\partial }-\rmi g
\not\! W) \phi \epsilon _R \qquad\,, \qquad \delta \lambda
=\frac{1}{4}\gamma ^{\mu \nu } F_{\mu \nu }\epsilon +\ft12\rmi \gamma _5
D  \epsilon  \,. \label{susy2} \eea
The covariant derivative on fermions is defined as $\nabla_\mu =  \partial _\mu  +\ft14 \omega _\mu
{}^{ab}(e)\gamma _{ab}$. The gravitino $U(1)$ connection  $A_\mu ^B$ plays an important role
in the gravitino transformations.
 \begin{eqnarray}
  A_\mu ^B =\frac{1}{2}\rmi\left[ \phi\hat{\partial} _\mu \phi^* -\phi^* \hat{\partial} _\mu \phi\right]
  +g W_\mu  \xi  \,.
  \label{AmuBinphi}
\end{eqnarray}
The cosmic string solutions to this theory found in \cite{dkvp} solve the Killing spinor equations \rf{susy1} - \rf{susy2} for some non-vanishing $\epsilon$. The metric ansatz in cylindrically symmetric form is
\begin{equation}
  \rmd s^2= -\rmd t^2 +\rmd z^2+\rmd r^2 + C^2(r) \rmd \theta ^2\,,
 \label{tentativemetric}
\end{equation}
where the plane of the string is parametrised by $r$ and $\theta $. We choose vierbein $e^1 =dr$ and $e^2 = C(r)d\theta$, which gives $\o_\theta^{12} = -C'(r)$ as the only non-vanishing spin connection component.

The Higgs field and gauge potential have the following form
\begin{equation}
\phi (r,\theta )\, = \, f(r)\,{\rm e}^{\rmi n \theta}\,, \qquad g W_\mu\, \rmd x^\mu = n\alpha (r) \,\rmd\theta \, .
 \label{stringhiggs}
\end{equation}
where $\theta$ is an azimuthal angle, and $f(r)$ is a real function that
outside the string core approaches the vacuum value $f^2=\xi$, for which
the $D$-term vanishes. One can solve for the profile functions $\alpha(r)$ and $C(r)$ explicitly in limiting cases, and one sees that the metric describes a spacetime with a conical deficit angle proportional to the Fayet-Iliopoulos constant $\xi$. A globally well-behaved spinor parameter is defined by
  $\epsilon _L(\theta) = \rme^{\mp\ft12\rmi\theta } \epsilon _{0L} \label{epsilontheta}$
where $\epsilon _{0L}$ is a constant spinor parameter satisfying $\gamma ^{12}\epsilon_{0}\, = \, \mp \rmi \gamma _5 \epsilon_{0}$. By demanding that the following condition holds
\begin{equation}
  1-C'(r) = \pm A_\theta ^B\,,
 \label{diffeqrho}
\end{equation}
one can then find solutions to the gravitino Killing spinor equation \rf{susy1}. As noted originally for three-dimensional supergravity \cite{3d}, the key to solving this Killing spinor equation in a conical spacetime is the $U(1)$ charge of the gravitino. This allows the singular spin connection term to be cancelled precisely because both the $U(1)$ charge and the deficit angle are set by the Fayet-Iliopoulos term $\xi$.

When the distance $r$ from the string core is large, the solution \rf{tentativemetric} takes the form of an asymptotically locally flat conical metric with an angular deficit angle due to the constant FI term $\xi$:
\be
  \rmd s^2= -\rmd t^2 +\rmd z^2+\rmd r^2 + r^2\left(1\mp n\xi \right)^2 \rmd \theta ^2\label{metricFAR}~, \ee
with the composite gauge field given by $A_\theta ^B= n\xi $.  Note that in the limit $r \rightarrow\infty$   the full supersymmetry is restored as $F_{\mu\nu}=0$, $D=0$,  $\partial_r \phi= \hat\partial_\theta \phi= 0$ and $R_{\mu\nu}{}^{ab}=0$, which corresponds to  the enhancement of supersymmetry away from the core of the string. 

In \cite{dkvp}, the string energy density was
defined using a Bogomol'nyi style argument. As the solution is
time-independent, the ansatz could be directly inserted into the action
with Gibbons-Hawking boundary terms included to give an energy
functional. The integral was then restricted to only run over directions
transverse to the string to ensure it produced a finite result. Using the
Bogomol'nyi method, this integral was then written in the following way
\begin{eqnarray}
{\cal \mu}_{\rm string}&=& \int \, \rmd r\rmd\theta\, C(r)\left\{  \,
 |(\hat{\partial}_r \phi  \, \pm \, {\rm i}C^{-1} \, \hat{\partial }_\theta ) \phi|^2 \, + \,
{\fft12}\left[ F_{12} \, \mp   D
\right]^2 \right\}  \,+\nonumber\\
&+& \int \rmd r \rmd\theta \,  \left[\partial _r\left( C'\pm
A_\theta\right) ^B\mp\partial _\theta A_r^B\right]-\left.\int
\rmd\theta\, C'\right|_{r=\infty }+\left.\int \rmd\theta\,
C'\right|_{r=0}~. \label{ebpsstring}
\end{eqnarray}
The condition arising from the gravitino Killing spinor equation~(\ref{diffeqrho}) implies that the first term in the second line in (\ref{ebpsstring}) vanishes. The first line vanishes by the remaining Killing spinor equations $\d \l = 0 = \d\chi_L$. The energy density is thus given by the difference between the boundary terms at $r=0$ and at $r =\infty$ \cite{dkvp}:
\begin{equation}
  \mu _{\rm string}=2\pi \left(\left.C'\right|_{r=0 }-\left.C'\right|_{r=\infty }\right)= \pm 2\pi n \xi \, ,
\label{energy}
\end{equation}
which agrees with the expected answer for a cosmic string solution \cite{rr}. It is interesting to note that as supersymmetry only fixed the metric function $C(r)$ up to a sign \cite{dkvp}, the energy also has a sign ambiguity. It was initially argued that the positive energy theorem implies that the negative energy solution should be ignored \cite{dkvp}. This is not correct, and the role of negative energy solution, i.e. the negative deficit angle solution, must be carefully reconsidered.

\section*{Negative deficit angle D-term strings}

A basic assumption in the proof of any positive energy theorem, which holds for all reasonable matter fields, is that the stress-energy tensor satisfies the dominant energy condition i.e. that for any timelike or null vector $u^a$, $-T^{~b}_{a} u^a $ is non-spacelike, which implies that  $T_{ab}u^a u^b \geq 0$.  It is straightforward to check that the supersymmetric Lagrangian \rf{bosonic2} satisfies the dominant energy condition, which can be conveniently restated as saying that matter energy density is non-negative in any orthonormal frame i.e. $T_{00} \geq 0$. For a general cylindrically symmetric spacetime, Comtet and Gibbons have shown that it is possible to find a useful rewriting of the metric in which it becomes clear that the sign of the string deficit angle $\d$ is completely determined by $T_{0 0}$: 
\begin{equation}
\delta \sim +\int_{\Sigma_2} T_{00} + ( \ldots )^2\,,
\end{equation}
where $\Sigma_2$ is a two-dimensional submanifold transverse to the string (we do not require the specific form of the terms in brackets). Hence, we see that it is only possible to have a solution with $\delta < 0$ if the Lagrangian violates the dominant energy condition. This implies that a $\d<0$ string is not a regular solution to the field equations derived from our Lagrangian \rf{bosonic2}, and therefore needs a source with negative $T_{0 0}$. Although the $\delta < 0$ solution is not known in closed form for small radius, the string ansatz we are using \eqref{tentativemetric} does not have a $g_{t t}$ component, and hence does not allow it to have horizons in the interior of the solution. This means that the defect, which requires the presence of a source, sweeps out a worldvolume over infinite time. In other words, the region of the solution that violates Einstein's equations is naked, which implies that no spacelike surface will be able to avoid it. Therefore no Cauchy surface exists in the $\d<0$ spacetime and so the positive energy theorem does not apply to it.

\section*{The positive energy theorem and semi-classical stability}

We shall now apply the standard Witten-Nester technique to prove the positive energy theorem for the positive deficit angle string backgrounds.
We begin by defining the generalised Witten-Nester 2-form \cite{WittenNester}:
\be
E^{\mu\nu} = \bar\eta \,\gamma^{\mu\nu\rho} \hat\nabla_\rho \eta~,
\ee
where we are using the supercovariant derivative defined by the gravitino supersymmetry transformation \rf{susy1}, and $\eta$ denotes a commuting spinor function that asymptotically tends to a background Killing spinor $\hat\nabla_\rho \eta =0$. We now define the Witten-Nester four-momentum as the integral of the dual of $E$
\begin{eqnarray}
P_\mu v^\mu =  \int_{\pt M} \ast E   = \frac{1}{2}\int_{\pt M} dS_{\mu\nu} E^{\mu\nu}  = \int_{M} d\Sigma_\nu \nabla_\mu E^{\mu\nu}~,
\end{eqnarray}
where $v^\mu = \bar\eta \g^\mu  \eta$. In the final equality we have
assumed that there are no internal boundaries, since the $\d>0$ D-term is regular \cite{dkvp}, and used
Gauss' law to write a volume integral. At this point, one should
understand that ${\pt M}$ is the two-dimensional boundary of an arbitrary
three-dimensional subsurface $M$. In order to evaluate the Witten-Nester total four-momentum explicitly for a particular solution, the surface charge integral must be regulated \cite{ght}. One must wrap the spatial worldvolume of the string such that ${\pt M} =\mathbf{R}_z \times S^1_{\theta} \rightarrow S^1_z \times S^1_{\theta}$, and then integrate out the $z$-contribution. The charge integral is then defined only over spatial directions transverse to the string, and it is formally the same as the equivalent three-dimensional expression \cite{3d}. By considering linearised perturbations that vanish asymptotically, one then sees that the surface integral form of the Witten-Nester total four-momentum becomes: 
\be
P_\mu v^\mu = \frac{1}{8}\int_{\pt M} dS_{\mu\nu} \varepsilon^{\mu\nu\rho\sigma}\varepsilon_{\d\a\b\s} \Delta\o_\rho^{\ul{\a\b}}e_{\ul \a}^\a e_{\ul \b}^\b\,\bar\eta \g^\d  \eta - \frac{1}{4}\int_{\pt M} dS_{\mu\nu} \varepsilon^{\mu\nu\rho\sigma} A_\rho^B \bar\eta \g_\s  \eta~. \label{lhsew}
\ee
The first term in \rf{lhsew}, which we shall denote $ {\cal P}_\m v^\m$, is Nester's expression for the gravitational four-momentum, where $\Delta\o_\rho^{\ul{\a\b}}$ is the difference of the spin connection with respect to the reference spacetime with $A_\rho^B =0$, i.e. Minkowski spacetime.  The second term in \rf{lhsew}, which we shall denote $J_\m^Rv^\m$, defines the R-charge of the string, i.e. the holonomy of composite gauge potential. 

In order to prove the positivity of the Witten-Nester four-momentum we now  turn to the volume integral expression. We want to show that an arbitrary on-shell perturbation of a supersymmetric
solution that vanishes asymptotically, but is otherwise unbounded,
contributes a positive amount to the total energy. As such, we shall not
wrap the spatial direction of the string worldvolume, such that $M$ is a
two-dimensional region, but consider the full three-dimensional volume
integral with $M=\mathbf{R}_r \times \mathbf{R}_z \times S^1_{\theta} $,
allowing for the most general perturbations. A lengthy calculation using
the standard manipulations \cite{WittenNester} then leads
to the following expression
 \be
P_\mu v^\mu = \int_{M} d\Sigma_\nu\left( \overline{\hat\nabla_\mu\eta}\g^{\mu\nu\rho}\hat\nabla_\rho\eta + \overline{\d\l} \g^\nu  \d\l + 2 \overline{\d\chi_L} \g^\nu  \d\chi_L + 2 \overline{\d\chi_R} \g^\nu  \d\chi_R\right)~, \label{rhs}
 \ee
where $\d\l$ and $\d\chi_{L,R}$ are the supersymmetry transformations \rf{susy2}, defined now with a commuting spinor parameter $\eta$. If we now choose $\Sigma$ to be an initial hypersurface with simple timelike norm and require that the spinors obey the generalised Witten condition $\g^j\hat\nabla_j\eta = 0$, we see that \rf{rhs} is then manifestly positive. As our spinors are Majorana, it is straightforward to show that the Killing vector $v^\m$ is non-spacelike and future directed, and thus that positivity of \rf{rhs} implies that energy in positive. 

As our choice of initial hypersurface $\Sigma$ was arbitrary, we can allow for arbitrary variations of it. This means that our expressions get promoted to fully covariant versions, and the Witten condition becomes $\g^\m\hat\nabla_\m\eta =0$. If we now use the covariant form of our result that $P_\mu \geq 0$ in conjunction with \rf{lhsew}, we reproduce the Bogomol'nyi bound for the $D$-term string:
 \be
P_\mu v^\mu = \overline\eta_0\left({\cal P}_\nu- J^R_\nu \right)\g^\n\eta_0 \geq0~. \label{ewbb}
\ee
Looking again at \rf{rhs}, we see that this inequality is saturated when the solution is supersymmetric, i.e. when $\d\l=\d\chi=\d\psi_\mu=0$, or equivalently the condition \rf{diffeqrho} holds. Here $\d\psi_i$ has been promoted to $\d\psi_\mu$ by allowing for arbitrary variations of the hypersurface $\Sigma$. It is possible to bring the Bogomol'nyi bound \rf{ewbb} into the more familiar form ${\cal P}_0 - Q^R \geq 0$ by taking the trace over the basis of spinors.

Our result proves that the positive deficit angle D-term string is stable against perturbations that vanish asymptotically, but are arbitrarily large in the bulk. An analogous result holds for point-like sources in three-dimensional supergravity \cite{3d}, however this is not sufficient to prove the stability of string solutions in four-dimensions, despite the fact that the linearised charge integrals agree. Instabilities in cylindrically symmetric spacetimes usually arise in the massive Kaluza-Klein tensor perturbations in the dimensionally reduced theory \cite{Gregory:1993vy}, and it is precisely this sector that is truncated in the three dimensional theory.

In order to complete the semi-classical proof, one must consider whether a non-perturbative quantum tunnelling effect could arise, i.e. a Coleman-de Luccia bounce solution \cite{Coleman:1980aw}. In other words, if the D-term string is viewed as a false vacuum state of the supergravity theory, is it possible to find a Euclidean bubble solution that would allow decay to a true vacuum state with lower energy? In \cite{Taylor-Robinson:1996fk} it was shown that such decay modes via bubble nucleation are inconsistent with ten- and eleven-dimensional supergravity theories, and the same result can be applied here. In short, if the putative true vacuum solution is required to asymptote to the original false vacuum solution, then it must also admit an asymptotically Killing spinor that is well-defined on the whole hypersurface. However, the positive energy theorem then implies that the energy of this solution can only be higher than that of the false vacuum, making it energetically unfavourable for the nucleation to take place. If the energy is equal to that of the false vacuum then the spinor must be globally Killing, which means the solution is no different from the false vacuum solution. In other words, no bubble is being nucleated. This completes the proof of the semi-classical stability of the positive deficit angle D-term string in four-dimensional $\cN=1$ supergravity with Fayet-Iliopoulos terms.

\subsection*{Acknowledgments}

We would like to thank Anne Davis, Jos{\'e} Edelstein, Dan Freedman, Gary Gibbons and Simon Ross for useful discussions. PS would also like to thank the organisers of the second workshop of the Marie Curie Research Training Network ForcesUniverse in Napoli, where this work was presented. This work is supported in part by the Federal Office for Scientific, Technical and Cultural Affairs through the ``Interuniversity Attraction Poles Programme -- Belgian Science Policy" P5/27 and by the European Community's Human Potential Programme under contract MRTN-CT-2004-005104 ``Constituents, fundamental forces and symmetries of the universe''.

\end{document}